\documentclass[twocolumn,prd,amssymb]{revtex4}
\usepackage{graphicx}

\def\Gammabol{{\stackrel{\circ}{\Gamma}}{}}

\def\Rbol{{\stackrel{\circ}{R}}{}}

\newcommand{\be}{\begin{equation}}
\newcommand{\ee}{\end{equation}}
\newcommand{\ba}{\begin{eqnarray}}
\newcommand{\ea}{\end{eqnarray}}

\begin{document}
\title{Selected Topics in Teleparallel Gravity}

\author{R. Aldrovandi}

\author{J. G. Pereira}

\author{K. H. Vu}

\affiliation{Instituto de F\'{\i}sica Te\'orica,
Universidade Estadual Paulista\\
Rua Pamplona 145, 01405-900 S\~ao Paulo SP, Brazil}

\begin{abstract}
Teleparallel gravity can be seen as a gauge theory for the translation group. As such,
its fundamental field is neither the tetrad nor the metric, but a gauge potential assuming
values in the Lie algebra of the translation group. This gauge character makes of
teleparallel gravity, despite its equivalence to general relativity,  a rather peculiar
theory. A first important point is that it does not rely on the universality of free fall,
and consequently does not require the equivalence principle to describe the gravitational
interaction. Another peculiarity is its similarity with Maxwell's theory, which allows an
Abelian nonintegrable phase factor approach, and consequently a global formulation for
gravitation. Application of these concepts to the motion of spinless particles, as well as
to the COW and gravitational Aharonov-Bohm effects are presented and discussed.
\end{abstract}

\maketitle

\section{Introduction}

Like the other fundamental interactions of nature, gravitation can be described by a
gauge theory \cite{PR}. The teleparallel equivalent of general relativity~\cite{tegra},
or teleparallel gravity for short \cite{obs}, can indeed be understood as a gauge theory
for the translation group. In this approach, the gravitational interaction is described by
a force similar to the Lorentz force equation of electrodynamics, with {\em torsion}
playing the role of force \cite{sp1}. 

On the other hand, due to the universality of free fall, it is also possible to describe
gravitation not as a {\em force}, but as a geometric {\em deformation} of flat
Minkowski spacetime. According to this point of view, a gravitational field produces a
{\em curvature} in spacetime, and its action on (structureless) particles is described
by letting them follow the geodesics of the curved spacetime. This is the approach of
general relativity, in which geometry replaces the concept of gravitational force, and
the trajectories are determined, not by force equations, but by geodesics. We notice in
passing that an immediate consequence of this dual description of gravitation is that
curvature and torsion might be related with the same degrees of freedom of the
gravitational field.

As a gauge theory for the translation group, which is an Abelian group, the teleparallel
formulation of gravity becomes in several aspects similar to the electromagnetic
Maxwell's theory. By exploring this analogy, as well as by using well known results of
electrodynamics, the basic purpose of this paper will be to study some specific
properties of teleparallel gravity. 

The first point to be examined refers to the weak equivalence principle, which
establishes the equality of inertial and gravitational masses. As is widely known, the
electromagnetic interaction is not universal and there exists no electromagnetic
equivalence principle. Nevertheless, Maxwell's theory, a gauge theory for the Abelian
group $U(1)$,  describes quite  consistently the electromagnetic interaction. Given the
analogy between electromagnetism and teleparallel gravity, in which the equations of
motion are not geodesics but force equations quite analogous to the electromagnetic
Lorentz force equation, the question then arises whether the gauge approach of
teleparallel gravity would also be able to describe the gravitational interaction in the
lack of universality, that is, in the absence of the weak equivalence principle. As we are
going to see, the answer to this question is positive: teleparallel gravity does not
require the validity of the equivalence principle to describe the gravitational
interaction. In fact, although the geometrical description of general relativity breaks
down, the gauge description of teleparallel gravity remains as a consistent theory in
the absence of universality \cite{wep}. It belongs, therefore, to a more general type
of theory.

A second point to be explored is the so called global formulation, which is an approach
based on the action of a nonintegrable phase factor. Relying on the well known phase
factor approach to Maxwell's theory \cite{wy}, a teleparallel nonintegrable phase factor
approach to gravitation will be developed, which represents the quantum mechanical version
of the classical gravitational Lorentz force of teleparallel gravity. As a first
application of this global approach, we consider the Colella, Overhauser, Werner (COW)
experiment \cite{cow}, which consists in using a neutron interferometer to observe the
quantum mechanical phase shift of neutrons caused by their interaction with Earth's
gravitational field. By considering the Newtonian limit, it is shown that the teleparallel
global formalism yields the correct quantum phase-shift predicted (as well as
experimentally verified) for the COW experiment. As a second application of the
teleparallel global approach, we obtain the quantum phase-shift produced by the coupling of
the particle's kinetic energy with the gravitomagnetic components of the translational
gauge potential \cite{global}. This effect is the gravitational analog of the usual
electromagnetic Aharonov-Bohm effect, and for this reason it will be called the
gravitational Aharonov-Bohm effect \cite{gab}. It is worthy mentioning that, as the
phase difference depends essentially on the energy, it applies equally to massive
and massless particles. For the sake of completeness, we begin by reviewing, in the next
section, the basic concepts related to teleparallel gravity. The equivalence principle is
recast in its language, and shown to be unnecessary. The global approach to gravitation is
then presented and applied to the two mentioned effects. 

\section{Fundamentals of Teleparallel Gravity}

Teleparallel gravity corresponds to a gauge theory of the translation group. According to
this model, to each point of spacetime there is attached a Minkowski tangent space, on
which the translation (gauge) group acts. We use the Greek alphabet $\mu, \nu, \rho, \dots
= 0, 1, 2, 3$ to denote spacetime indices and the Latin alphabet $a, b, c, \dots = 0, 1, 2,
3$ to denote anholonomic indices related to the tangent Minkowski spaces, whose metric is
chosen to be $\eta_{a b} = {\rm diag} (+1, -1, -1, -1)$. As a gauge theory for
translations, the fundamental field of teleparallel gravity is the translational gauge
potential $B^a{}_\mu$, a 1-form assuming values in the Lie algebra of the translation
group:
\be
B_\mu = B^a{}_\mu \, P_a,
\ee
with $P_a = \partial_a$ the generators of infinitesimal translations. Under a local
translation of the tangent space coordinates $\delta x^a = \epsilon^a(x)$, the gauge
potential transforms according to
\be
B^{\prime a}{}_\mu = B^a{}_\mu - \partial_\mu \epsilon^a.
\label{btrans}
\ee
It appears naturally as the nontrivial part of the tetrad field $h^{a}{}_{\mu}$:
\be
h^a{}_\mu = \partial_\mu x^a + B^a{}_\mu.
\label{tetrada}
\ee
Notice that, whereas the tangent space indices are raised and lowered with the Minkowski
metric $\eta_{a b}$, the spacetime indices are raised and lowered with the spacetime
metric
\be
g_{\mu \nu} = \eta_{a b} \; h^a{}_\mu \; h^b{}_\nu.
\label{gmn}
\ee

The above tetrad gives rise to the so called Weit\-zen\-b\"ock connection
\begin{equation}
\Gamma^{\rho}{}_{\mu\nu} = h_{a}{}^{\rho}\partial_{\nu}h^{a}{}_{\mu},
\label{carco}
\end{equation}
which introduces the distant parallelism in the four-dimensional spacetime manifold. It
is a connection which presents torsion, but no curvature. Its torsion,
\begin{equation}
T^{\rho}{}_{\mu\nu} = \Gamma^{\rho}{}_{\nu\mu} - 
\Gamma^{\rho}{}_{\mu\nu},
\label{tor}
\end{equation}
is related to the translational gauge field strength $F^a{}_{\mu \nu}$ by
\begin{equation}
F^a{}_{\mu \nu} \equiv \partial_\mu B^a{}_{\nu} - \partial_\nu B^a{}_{\mu} =
h^a{}_\rho \; T^\rho{}_{\mu \nu}.
\label{gfs}
\end{equation}
The Weitzenb\"ock connection can be decomposed as
\begin{equation}
\Gamma^{\rho}{}_{\mu\nu} = \Gammabol^{\rho}{}_{\mu\nu} 
+ K^{\rho}{}_{\mu\nu},
\label{rela}
\end{equation}
where $\Gammabol^{\rho}{}_{\mu\nu}$ is the Christoffel connection constructed from the
spacetime metric $g_{\mu\nu}$, and
\begin{equation}
K^{\rho}{}_{\mu \nu} = \frac{1}{2} \left( 
T_{\mu}{}^{\rho}{}_{\nu} + T_{\nu}{}^{\rho}{}_{\mu} 
- T^{\rho}{}_{\mu \nu} \right)
\label{contorsion}
\end{equation}
is the contortion tensor. It is important to remark that curvature and torsion are
considered as properties of a connection, not of spacetime \cite{livro}. Notice, for
example, that the Christoffel and the Weitzenb\"ock connections are defined on the very
same spacetime manifold.

The Lagrangian of the teleparallel equivalent of general relativity is \cite{sp1}
\begin{equation}
{\mathcal L} \equiv {\mathcal L}_G + {\mathcal L}_M =
\frac{c^{4} h}{16\pi G} \, S^{\rho\mu\nu}\,T_{\rho\mu\nu} + {\mathcal L}_M,
\label{gala}
\end{equation}
where $h = {\rm det}(h^{a}{}_{\mu})$, ${\mathcal L}_M$ is the Lagrangian of a source
field, and
\begin{equation}
S^{\rho\mu\nu} = - S^{\rho\nu\mu} = \frac{1}{2} 
\left[ K^{\mu\nu\rho} - g^{\rho\nu}\,T^{\sigma\mu}{}_{\sigma} 
+ g^{\rho\mu}\,T^{\sigma\nu}{}_{\sigma} \right]
\label{S}
\end{equation}
is a tensor written in terms of the Weitzenb\"ock connection only. Performing a
variation with respect to the gauge potential, we find the teleparallel version of the
gravitational field equation \cite{sp2},
\begin{equation}
\partial_\sigma(h S_\lambda{}^{\rho \sigma}) -
\frac{4 \pi G}{c^4} \, (h t_\lambda{}^\rho) =
\frac{4 \pi G}{c^4} \, (h {\mathcal T}_\lambda{}^\rho),
\label{eqs1}
\end{equation}
where
\begin{equation}
h \, t_\lambda{}^\rho = \frac{c^4 h}{4 \pi G} \, S_{\mu}{}^{\rho \nu}
\,\Gamma^\mu{}_{\nu\lambda} -\, \delta_\lambda{}^\rho \, {\mathcal L}_G
\label{emt}
\end{equation}
is the energy-momentum pseudotensor of the gravitational field, and
${\mathcal T}_\lambda{}^\rho = {\mathcal T}_a{}^\rho \, h^a{}_\lambda$ is the
energy-momentum tensor of the source field, with
\be
h \, {\mathcal T}_a{}^\rho = -\,
\frac{\delta {\mathcal L}_M}{\delta B^a{}_\rho} \equiv -\,
\frac{\delta {\mathcal L}_M}{\delta h^a{}_\rho}.
\ee
A solution of the gravitational field
equation (\ref{eqs1}) is an explicit form of the gravitational gauge potential
$B^a{}_\mu$.

When the weak equivalence principle is assumed to be true, teleparallel gravity turns
out to be equivalent to general relativity. In fact, up to a divergence, the Lagrangian
(\ref{gala}) is found to be equivalent to the Einstein-Hilbert Lagrangian of general
relativity, and the teleparallel field equation (\ref{eqs1}) is found to coincide with
Einstein's equation
\begin{equation}
\Rbol_\lambda{}^\rho - \frac{1}{2} \, \delta_\lambda{}^\rho \, \Rbol =
\frac{8 \pi G}{c^4} \, {\mathcal T}_\lambda{}^\rho,
\label{einsteinbol}
\end{equation}
with $\Rbol_\lambda{}^\rho$ and $\Rbol$ respectively the Ricci and the scalar curvature of
the Christoffel connection.

\section{Gravitation and the Weak Equivalence Principle}

Let us begin by making it clear that, in spite of many controversies related with the
equivalence principle \cite{pequi}, it is not our intention here to question its validity,
but simply to verify whether teleparallel gravity requires it or not to describe the
gravitational interaction. This will be done by supposing that the gravitational mass $m_g$
and the inertial mass $m_i$ do not coincide, and then by making a comparative study of the
force equation of teleparallel gravity with the geodesic equation of general relativity.

\subsection{Teleparallel Gravity: Force Equation}

Let us then consider, in the context of teleparallel gravity, the motion of a
spinless particle in a gravitational field $B^{a}{}_{\mu}$, supposing however that the
gravitational and the inertial masses do not coincide. Analogously to the
electromagnetic case \cite{landau}, the action integral is written in the form
\be
S = \int_{a}^{b} \left[ -\, m_i \, c \, d\sigma -
m_g \, c \, B^{a}{}_{\mu} \, u_{a} \, dx^{\mu} \right],
\label{acao1}
\ee
where $d\sigma = (\eta_{a b} dx^a dx^b)^{1/2}$ is the Minkowski tangent-space invariant
interval, and $u^a$ is the particle four-velocity seen from the tetrad frame,
necessarily anholonomic when expressed in terms of the {\it spacetime} line element
$ds$. The first term of the action (\ref{acao1}) represents the action of a free
particle, and the second the (minimal) coupling of the particle with the gravitational
field. Variation of the action (\ref{acao1}) yields the equation of motion \cite{wep}
\be
\left( \partial_\mu x^a +
\frac{m_g}{m_i} \; B^a{}_\mu \right) \frac{d u_a}{d s} =
\frac{m_g}{m_i} \; F^a{}_{\mu \rho} \; u_a \, u^\rho,
\label{eqmot2}
\ee
where $F^a{}_{\mu \rho}$ is the gravitational field strength defined in
Eq.~(\ref{gfs}), and
\be
u^\mu = \frac{d x^\mu}{ds} \equiv h^\mu{}_a \, u^a
\label{ust}
\ee
is the holonomic four-velocity, with $ds$ = $(g_{\mu \nu} dx^\mu dx^\nu)^{1/2}$
the Riemannian spacetime invariant interval. Equation (\ref{eqmot2}) is the force
equation governing the motion of the particle, in which the teleparallel field strength
$F^a{}_{\mu \rho}$ (that is, the  Weitzenb\"ock torsion) plays the role of
gravitational force. Similarly to the electromagnetic Lorentz force, which depends on
the relation $e/m_i$, with $e$ the electric charge of the particle, the gravitational
force depends explicitly on the relation ${m_g}/{m_i}$ of the particle.

We see from the above equations that, even in the absence of the weak equivalence
principle, teleparallel gravity is able to describe the motion of a particle with $m_g
\neq m_i$. The crucial point is to observe that, although the equation of motion
depends explicitly on the relation  ${m_g}/{m_i}$ of the particle, neither $B^a{}_\mu$
nor $F^a{}_{\rho \mu}$ depends on this relation. This means essentially that the
teleparallel field equation (\ref{eqs1}) can be consistently solved for the
gravitational potential $B^a{}_\mu$, which can then be used to write down the equation
of motion (\ref{eqmot2}), independently of the validity or not of the weak equivalence
principle. The gauge potential $B^a{}_\mu$, therefore, may be considered as the most
fundamental field representing gravitation. As we are going to see next, this is not
the case of general relativity, in which to keep the equations of motion given by
geodesics, the gravitational field (metric tensor) must necessarily depend on the relation 
${m_g}/{m_i}$ of the particle, rendering thus the theory inconsistent when $m_g \neq m_i$.

\subsection{General Relativity: Geodesics}

According to teleparallel gravity, even when $m_g \neq m_i$, the tetrad is still given by
(\ref{tetrada}), and the spacetime indices are raised and lowered with the metric
(\ref{gmn}). Then, by using the relation (\ref{gfs}), as well as the identity
\be
T^\lambda{}_{\mu \rho} \, u_\lambda \, u^\rho = -\, K^\lambda{}_{\mu \rho}
\, u_\lambda \, u^\rho,
\ee
the force equation (\ref{eqmot2}) can be rewritten in the form
\be
\frac{d u_\mu}{ds} - \Gammabol^\lambda{}_{\mu \rho} \, u_\lambda \, u^\rho =
\left(\frac{m_g - m_i}{m_g} \right) \partial_\mu x^a \, \frac{d u_a}{d s},
\label{eqmot6}
\ee
where use has been made also of the relation (\ref{rela}). Notice that the violation of
the weak equivalence principle produces a deviation from the geodesic motion, which is
proportional to the difference between the gravitational and inertial masses. Notice
furthermore that, due to the assumed non-universality of free fall, it is not possible
to find a local coordinate system in which the gravitational effects are absent.

Now, as already said, when the weak equivalence principle is assumed to be true, the
teleparallel field equation (\ref{eqs1}) is equivalent to Einstein's equation
(\ref{einsteinbol}). Accordingly, when $m_g = m_i$, the equation of motion (\ref{eqmot2})
reduces to the geodesic equation of general relativity, as can be seen from its equivalent
form (\ref{eqmot6}). However, in the absence of the weak equivalence principle, it is not a
geodesic equation. This means that the equation of motion (\ref{eqmot2}) does not
comply with the geometric description of general relativity, according to which all
trajectories must be given by genuine geodesic equations. In order to comply with the
foundations of general relativity, it is necessary to incorporate the particle
properties into the geometry. This can be achieved by assuming, instead of the tetrad
(\ref{tetrada}) of teleparallel gravity, the new tetrad
\be
\bar{h}^a{}_\mu = \partial_\mu x^a + \frac{m_g}{m_i} \; B^a{}_\mu,
\label{tetrada2}
\ee
which takes into account the characteristic $m_g/m_i$ of the particle under consideration.
This tetrad defines a new spacetime metric tensor
\be
\bar{g}_{\mu \nu} = \eta_{a b} \; \bar{h}^a{}_\mu \; \bar{h}^b{}_\nu,
\label{gmn2}
\ee
in terms of which the corresponding spacetime invariant interval is
\be
d\bar{s}^2 = \bar{g}_{\mu \nu} \, dx^\mu dx^\nu.
\ee
By noticing that in this case the relation between the gravitational field strength and
torsion becomes
\be
\frac{m_g}{m_i} \; F^a{}_{\mu \rho} = \bar{h}^a{}_\lambda \, \bar{T}^\lambda{}_{\mu \rho},
\label{fstor2}
\ee
it is an easy task to verify that, for a fixed relation $m_g/m_i$, the equation of motion
(\ref{eqmot2}) is equivalent to the true geodesic equation
\be
\frac{d \bar{u}_\mu}{d\bar{s}} - {\bar{\Gamma}}{}^\lambda{}_{\mu \rho}
\, \bar{u}_\lambda \, \bar{u}^\rho = 0,
\label{eqmot7}
\ee
where $\bar{u}_\mu \equiv d x_\mu/d \bar{s} = \bar{h}^a{}_\mu u_a$, and
$\bar{\Gamma}{}^{\rho}{}_{\mu\nu}$ is the Christoffel connection of the metric
$\bar{g}_{\mu \nu}$.
However, the price for imposing a geodesic equation of motion to describe a
non-universal interaction is that the gravitational theory becomes inconsistent. In
fact, the solution of the corresponding Einstein's field equation
\be
\bar{R}_{\mu \nu} - \frac{1}{2} \, \bar{g}_{\mu \nu} \bar{R} =
\frac{8 \pi G}{c^4} \, \bar{\mathcal T}_{\mu \nu},
\label{e2}
\ee
which is not equivalent to any teleparallel field equation, would in this case depend
on the relation $m_g/m_i$ of the test particle, which renders the theory inconsistent
in the sense that test particles with different relations $m_g/m_i$ would require
connections with different curvatures to keep all equations of motion given by
geodesics. Of course, as a true field, the gravitational field cannot depend on any
test particle properties.

\section{Global Formulation of Gravitation}

The basic conclusion of the previous section is that the fundamental field describing
gravitation is neither the tetrad nor the metric, but the translational gauge potential
$B^a{}_\mu$. Using this fact, and the similarity of teleparallel gravity with Maxwell's
theory, we are going to introduce now a teleparallel nonintegrable phase factor, in
terms of which a global formulation for gravitation will be developed.

\subsection{Nonintegrable Phase Factor}

As is well known, in addition to the usual {\em differential} formalism,
electromagnetism presents also a {\em global} formulation in terms of a nonintegrable
phase factor \cite{wy}. According to this approach, electromagnetism can be considered
as the gauge invariant effect of a nonintegrable (path-dependent) phase factor. For a
particle with electric charge $e$ traveling from an initial point ${\sf P}$ to a final
point ${\sf Q}$, the phase factor is given by
\be
\Phi_e({\sf P}|{\sf Q}) = \exp \left[\frac{i e}{\hbar c} \int_{\sf P}^{\sf Q}
A_\mu \, dx^\mu \right],
\ee
where $A_\mu$ is the electromagnetic gauge potential. In the classical (non-quantum)
limit, the action of this nonintegrable phase factor on a particle wave-function yields
the same results as those obtained from the Lorentz force equation
\be
\frac{d u^a}{ds} = \frac{e}{m_i c^2} \, F^a{}_b \, u^b.
\ee
In this sense, the phase-factor approach can be considered as the {\em quantum}
generalization of the {\em classical} Lorentz force equation. It is actually
more general, as it can be used both on simply-connected and on multiply-connected
domains. Its use is mandatory, for example, to describe the Aharonov-Bohm effect, a
quantum phenomenon taking place in a multiply-connected space \cite{ab}.

Now, in the teleparallel approach to gravitation, the fundamental field describing
gravitation is the translational gauge potential $B^a{}_\mu$. Like $A_\mu$, it is an
Abelian gauge potential. Thus, in analogy with electromagnetism, $B^a{}_\mu$ can be
used to construct a global formulation for gravitation. To start with, let us notice
that the electromagnetic phase factor $\Phi_e({\sf P}|{\sf Q})$ is of the form
\be
\Phi_e({\sf P}|{\sf Q}) = \exp \left[\frac{i}{\hbar} \, S_e \right],
\ee
where $S_e$ is the action integral describing the interaction of the charged particle
with the electromagnetic field. Now, in teleparallel gravity, the action integral
describing the interaction of a particle of mass $m_g$ with gravitation, according to
Eq.~(\ref{acao1}), is given by
\be
S_g = \int_{\sf P}^{\sf Q} m_g \, c \, B^a{}_\mu \, u_a \, dx^\mu.
\ee
Therefore, the corresponding gravitational nonintegrable phase factor turns out
to be
\be
\Phi_g({\sf P}|{\sf Q}) = \exp \left[\frac{i m_g c}{\hbar} \int_{\sf P}^{\sf Q}
B^a{}_\mu \, u_a \, dx^\mu \right].
\label{npf}
\ee
Similarly to the electromagnetic phase factor, it represents the {\em quantum}
mechanical law that replaces the {\em classical} gravitational Lorentz force
equation (\ref{eqmot2}).

\subsection{The COW Experiment}

As a first application of the gravitational nonintegrable phase factor (\ref{npf}), we
consider the COW experiment \cite{cow}. It consists in using a neutron interferometer
to observe the quantum mechanical phase shift of neutrons caused by their interaction
with Earth's gravitational field, which is usually assumed to be Newtonian.
Furthermore, as the experience is performed with thermal neutrons, it is possible to
use the small velocity approximation. In this case, the gravitational phase factor
(\ref{npf}) becomes
\be
\Phi_g({\sf P}|{\sf Q}) = \exp \left[\frac{i m_g c^2}{\hbar} \int_{\sf P}^{\sf Q}
\, B_{00} \, dt \right],
\label{npf2}
\ee
where we have used that $u^0 = \gamma \simeq 1$ for the thermal neutrons. In the Newtonian
approximation, we can set $c^2 B_{00} \equiv \phi = g \, z$, with $\phi$ the (homogeneous)
Earth Newtonian potential \cite{global}. In this expression, $g$ is the grav\-itational
acceleration, assumed not to change significantly in the region of the experience, and $z$
is the distance from Earth taken from some reference point. Consequently, the phase factor
can be rewritten in the form
\be
\Phi_g({\sf P}|{\sf Q}) = \exp \left[\frac{i m_g g}{\hbar} \int_{\sf P}^{\sf Q}
z(t) \, dt \right] \equiv \exp i \varphi.
\label{npf3}
\ee
\begin{figure}[ht]
\begin{center}
\includegraphics[height=5.5cm,width=6.5cm]{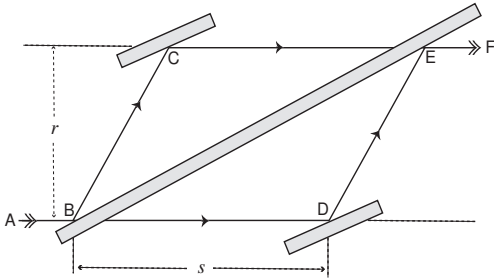}
\end{center}
\vspace{-40pt}
\caption{Schematic illustration of the COW neutron interferometer.}
\label{fig1}
\end{figure}

Let us now compute the phase $\varphi$ through the two trajectories of
Figure~\ref{fig1}. As the phase contribution along the segments {\sf DE} and {\sf BC}
are equal, they cancel out and do not contribute to the phase. Assuming that the
segment {\sf BD} is at $z=0$, we obtain for the trajectory {\sf BDE}:
\be
\varphi_{\sf BDE} = \frac{m_g g}{\hbar} \int_{\sf D}^{\sf E} z(t) \, dt.
\ee
For the trajectory {\sf BCE}, we have
\be
\varphi_{\sf BCE} = \frac{m_g g}{\hbar} \int_{\sf B}^{\sf C} z(t) \, dt + \frac{m_g
g r}{\hbar} \int_{\sf C}^{\sf E} dt.
\ee
Therefore, we get
\be
\Delta \varphi \equiv \varphi_{\sf BCE} - \varphi_{\sf BDE} =
\frac{m_g g r}{\hbar} \int_{\sf C}^{\sf E} dt.
\ee
Since the neutron velocity is constant along the segment {\sf CE}, we have
\be
\int_{\sf C}^{\sf E} dt \equiv \frac{s}{v} = \frac{s m_i \lambda}{h}\,\,\, ,
\ee
where $s$ is the length of the segment {\sf CE}, and $\lambda = h/(m_i v)$ is the
de Broglie wavelength associated with the neutron. The gravitationally induced phase
difference predicted for the COW experience is then found to be \cite{global}
\be
\Delta \varphi = s \, \frac{2 \pi g r \lambda m_i^2}{h^2}
\left(\frac{m_g}{m_i} \right).
\ee
When the gravitational and inertial masses are assumed to coincide, the phase shift
becomes
\be
\Delta \varphi =  s \, \frac{2 \pi g r \lambda m^2}{h^2},
\ee
which is exactly the result obtained for the COW experiment \cite{cow}.

\subsection{Gravitational Aharonov-Bohm Effect}

As a second application we use the phase factor (\ref{npf}) to study the gravitational
analog of the Aharonov-Bohm effect \cite{gab}. The usual (electromagnetic) Aharonov-Bohm
effect consists in a shift, by a constant amount, of the electron interferometry wave
pattern, in a region where there is no magnetic field, but there is a nontrivial gauge
potential $A_i$. Analogously, the gravitational Aharonov-Bohm effect will consist in a
similar shift of the same wave pattern, but produced by the presence of a gravitational
gauge potential $B_{0 i}$. Phenomenologically, this kind of effect might be present
near a massive rapidly rotating source, like a neutron star, for example. Of course,
differently from an ideal apparatus, in a real situation the gravitational field cannot
be completely eliminated, and consequently the gravitational Aharonov-Bohm effect
should be added to the other effects also causing a phase change.

\begin{figure}[ht]
\begin{center}
\includegraphics[height=4.0cm,width=7.5cm]{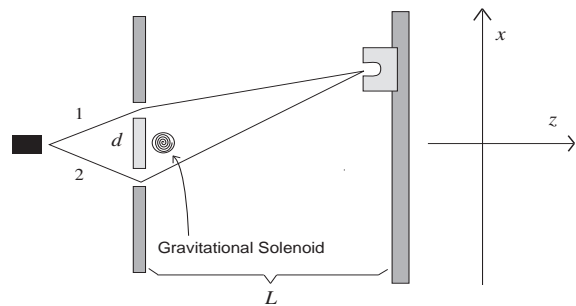}
\end{center}
\caption{Schematic illustration of the gravitational Aharonov-Bohm electron
interferometer.}
\label{fig2}
\end{figure}

Let us consider first the case in which there is no external field at all. If the
electrons are emitted with a characteristic momentum $p$, then its wavefunction has the
de Broglie wavelength $\lambda = h/p$. Denoting by $L$ the distance between slit and
screen (see Figure~\ref{fig2}), and by $d$ the distance between the two holes, when the
conditions $L \gg \lambda$, $L \gg x$ and $L \gg d$ are satisfied, the phase difference
at a distance $x$ from the central point of the screen is given by
\be
\delta^0 \varphi(x) = \frac{2 \pi x d}{L \lambda}.
\label{wpat}
\ee
This expression defines the wave pattern on the screen.

We consider now the ideal case in which a kind of infinite ``gravitational solenoid''
produces a purely static gravitomagnetic field flux concentrated in its interior. In the
ideal situation, the gravitational field outside the solenoid vanishes completely, but
there is a nontrivial gauge potential $B_{0 i}$. When we let the electrons to move
outside the solenoid, phase factors corresponding to paths lying on one side of the
solenoid will interfere with phase factors corresponding to paths lying on the other
side, which will produce an additional phase shift at the screen. Let us then calculate
this additional phase shift. The gravitational phase factor (\ref{npf}) for the
physical situation described above is
\be
\Phi_g({\sf P}|{\sf Q}) = \exp \left[- \frac{i m_g c}{\hbar} \int_{\sf P}^{\sf Q}
u^0 \vec{B}_0 \cdot d\vec{r} \right],
\ee
where $\vec{B}_0$ is the vector with components $B_0{}^i = -\, B_{0i}$. Since
$u^0 = \gamma \equiv [1 - ({v^2}/{c^2}) ]^{-1/2}$, and considering that the
electron velocity $v$ is constant, the phase difference at the screen will be
\be
\delta \varphi \equiv \varphi_2 -\varphi_1 =
\frac{\gamma m_g c}{\hbar} \oint \vec{B}_0 \cdot d\vec{r}.
\ee
Since the integral
\be
\oint \vec{B}_0 \cdot d\vec{r} = \oint (\vec{\nabla} \times \vec{B}_0)
\cdot d\vec{\sigma} = \oint \vec{H} \cdot d\vec{\sigma} \equiv \Omega
\ee
represents the flux $\Omega$ of the gravitomagnetic field $\vec{H} = \vec{\nabla} \times
\vec{B}_0$ inside the solenoid, the phase shift can be written in the form
\be
\delta \varphi = \frac{{\mathcal E} \, \Omega}{\hbar \, c} \left(\frac{m_g}{m_i}
\right),
\label{gabe1}
\ee
where ${\mathcal E} = \gamma m_i c^2$ is the electron kinetic energy. When the
gravitational and inertial masses are assumed to coincide, the phase shift becomes
\be
\delta \varphi = \frac{{\mathcal E} \, \Omega}{\hbar \, c}.
\label{gabe2}
\ee

Expression (\ref{gabe2}) gives the phase difference produced by the interaction of the
particle's kinetic energy with a gauge potential, which gives rise to the gravitational
Aharonov-Bohm effect. As this phase difference depends on the energy, it applies
equally to massive and massless particles. There is a difference, however: whereas for
a massive particle it is a genuine quantum effect, for massless particles, due to the
their intrinsic wave character, it can be considered as a classical effect. In fact,
for ${\mathcal E}=\hbar \omega$, Eq.~(\ref{gabe2}) becomes
\be
\delta \varphi =
\frac{\omega \, \Omega}{c},
\label{gabeb}
\ee
and we see that, in this case, the phase difference does not depend on the Planck's
constant. It is important to remark that, like the electromagnetic case, the phase
difference is independent of the position $x$ on the screen, and consequently the whole
wave pattern defined by (\ref{wpat}) will be shifted by a constant amount.

\section{Final Remarks}

In Einstein's general relativity, a theory fundamentally based on the universality of
free fall (or on the weak equivalence principle), geometry replaces the concept of
gravitational force. This theory has been confirmed by all experimental tests at the
classical level \cite{exp}, but any violation of the principle would lead to its ruin. We
notice in passing that the non-universality of the electromagnetic interaction is the
reason why there is no geometric description, in the sense of general relativity, for
electromagnetism.

On the other hand, the teleparallel equivalent of general relativity does not geometrize
the interaction, but shows gravitation as a gauge force quite analogous to the Lorentz
force of electrodynamics. It is  able to describe the gravitational interaction in the
absence of universality just as Maxwell's gauge theory is able to describe the non-universal
electromagnetic interaction. In spite of the equivalence with general relativity
\cite{always}, it can  be considered as a more fundamental theory as it dispenses with one
assumption. Notice in this connection that the equivalence principle is frequently said to
preclude the definition of a local energy-momentum density for the gravitational
field \cite{gravitation}. Although this is a true assertion in the context of general
relativity, it has already been shown that a tensorial expression for the gravitational
energy-momentum density is possible in the context of teleparallel gravity \cite{sp2},
which shows the consistency of the results.

Now, at the quantum level, deep conceptual changes occur with respect to classical
gravity, the most important being the fact that gravitation seems to be no more
universal \cite{nonuni}. In fact, at this level, the phase of the particle wavefunction
acquires a fundamental status, and turns out to depend on the particle mass (in the COW
effect, obtained in the non-relativistic limit), or on the relativistic kinetic energy
(in the gravitational Aharonov-Bohm effect). Although in the specific case of the
 COW experiment the phase shift can be made independent of the mass by introducing a kind
of quantum equivalence principle \cite{lammerzahl}, the basic difficulty remains that
different versions of this quantum principle would be necessary for different phenomena.
Since teleparallel gravity is able to describe gravitation independently of the validity or
not of the equivalence principle \cite{wep}, it will not require a quantum version of this
principle to deal with gravitationally induced quantum effects, and can be considered
as providing a much more appropriate and consistent approach to study such effects. 

Relying on the above arguments, we can say that the fundamental field describing
gravitation is neither the tetrad nor the metric, but the translational gauge potential
$B^a{}_\mu$. Metric is no more a fundamental, but a derived quantity. This point can have
important consequences for both classical and quantum gravity. Gravitational waves should
be seen as $B$ waves and not as metric waves. Quantization of the gravitational field
should be carried out on $B^a{}_\mu$ and not on the metric. Another consequence refers to
a fundamental problem of quantum gravity, namely, the conceptual difficulty of reconciling
{\it local} general relativity with {\it non-local} quantum mechanics, or of reconciling
the local character of the equivalence principle with the non-local character of the
uncertainty principle \cite{qep}. As teleparallel gravity can be formulated independently
of the equivalence principle, the quantization of the gravitational field may possibly
appear more consistent if considered in the teleparallel picture.

\begin{acknowledgments}
The authors thank J. M. Nester for useful discussions. They also thank FAPESP-Brazil,
CAPES-Brazil, and CNPq-Brazil for partial financial support.
\end{acknowledgments}

\end{document}